# Single Molecule Mixture: A Concept in Polymer Science

Yu Tang*

**Abstract:** In theory, there exist two extreme forms of substances: pure form and single-molecule mixture form. Single-molecule mixture form contains a mixture of molecules that have molecularly different structures. This elusive form has not yet been explored. Herein, we report a study of single molecule mixture state by a combination of model construction and mathematical analysis, and a series of interesting results were obtained. These results provide theoretical evidence that single-molecule mixture state may indeed exist in realistic synthetic or natural polymer system.

Realistic substances always contain a mixture of molecules, in theory, there exist two extreme forms of substances: pure form and single-molecule mixture form, as is shown in Figure 1. Pure form contains only one kind of molecule. Single-molecule mixture form, on the other hand, contains a mixture of molecules that have molecularly different structures. It is noted that this elusive form of molecules has not yet been explored. A very interesting question is whether single molecule mixture form substance exists in nature or synthetic systems. Herein, we report a theoretical exploration of single molecule mixture form and provide theoretical evidence that this form of molecules may indeed distributed in nature or synthetic polymer systems.

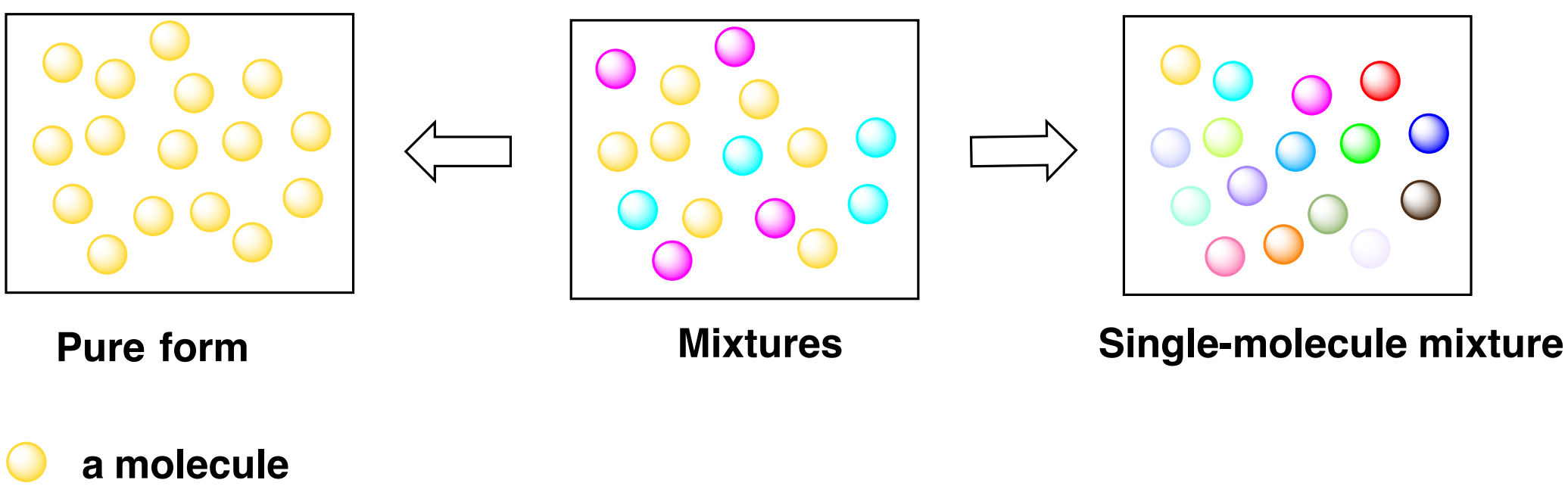

**Figure 1.** Two extreme forms of substances: pure form and single-molecule mixture form.

To start our exploration, a model polymer molecule system was first constructed,

as illustrated in Figure 2. This system contains $a$ substitutional sites, each site is randomly substituted by one of the two possible alkyl groups, $R^1$ or $R^2$, thus, the overall number of the structural isomers of in this structural space are $2^a$. As $a$ increases, the overall number of isomers increases exponentially. For example, if $a$ = 200, then the overall number of isomers of this system is $2^{200}$, namely, $1.60693804\times10^{60}$.

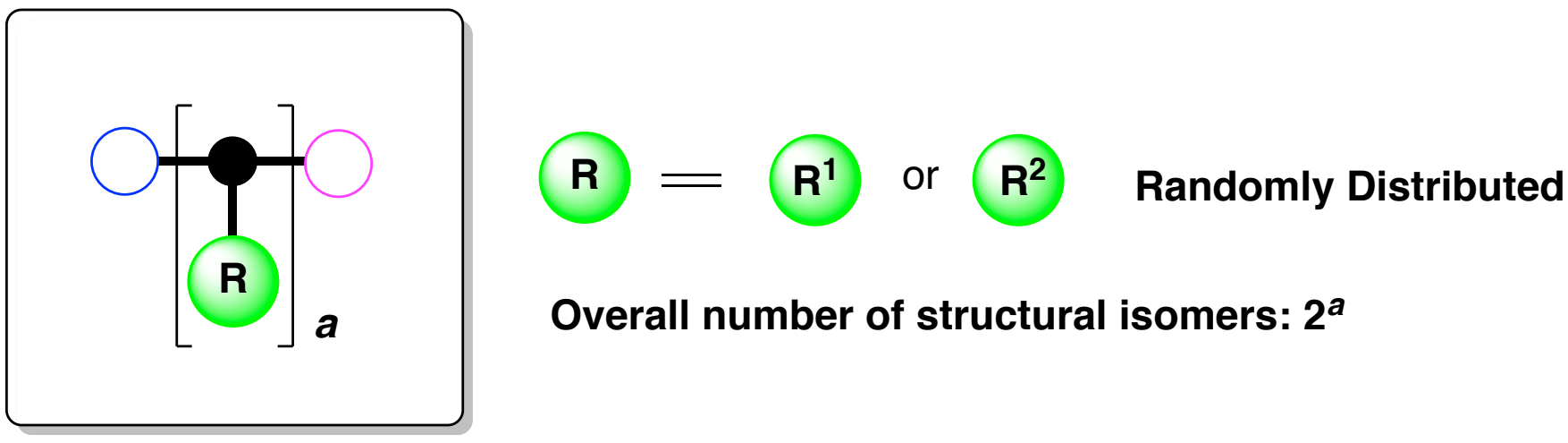

**Figure 2.** A model polymer molecule system

A very interesting question occurs: if we take a given number ($n$) of structures from the above structural space (overall number of structures = $m$) individually and randomly (which means that the same structure may be taken more than once), how to calculate the probability ($P$) of getting single-molecule mixture (in that each structure we take are different)?

This problem can be solved through probability calculation.

$$P = \frac{C_{m-n}^{n}}{C_{m}^{n}} = \frac{(m-n)(m-n-1)(m-n-2)\cdots\cdots(m-n-n)}{m(m-1)(m-2)\cdots\cdots(m-n)}$$

$$= (1-\frac{n}{m})(1-\frac{n}{m-1})\ (1-\frac{n}{m-2})\cdots\cdots(1-\frac{n}{m-n})$$

If $m > n^2$, then, $P > (1-\frac{n}{m-n})^{n} > 1-n\times\frac{n}{m-n}$

If $n^2>m>2n$, then $P < (1-\frac{n}{m})^{n}$

In this case, we choose $n = N_A$ ($6.02\times10^{23}$) as the given number of molecules taken from the structural space, and $a$ = 200 as the number of substitutional sites in the polymer model system (thus $m = 2^{200} = 1.60693804\times10^{60}$), then the probability ($P$) of getting single-molecule mixture is:

$$P > (1-\frac{n}{m-n})^{n} > 1-n\times\frac{n}{m-n} \approx 1 - 2.26\times10^{-13}$$

This result reveals that if we individually and randomly take $n = N_A$ ($6.02\times10^{23}$) molecules from this model structure space ($a$ = 200, $m = 1.60693804\times10^{60}$), the probabilities ($P$) of gettiing single-molecule mixtures close to 1.

It is noted that in many realistic synthetic polymer systems and natural biopolymer systems, partial random substitution of the parent molecule by a few substituent groups frequently occurred, such as in polymer bromination process[1,2] (random partial substitution of the hydrogen atom by a bromine atom) and many other post-polymerization modification process[3-10], and DNA methylation process[11,12] (random partial substitution of the hydrogen atom by a methyl group) and protein methylation process(random partial substitution of the hydrogen atom by a methyl group)[13,14], as illustrated in Figure 3 and Figure S1. This substitution process will generate a large number of isomers with equal probabilities. In order to calculate the exact number of potential isomers, we constructed two additional model systems.

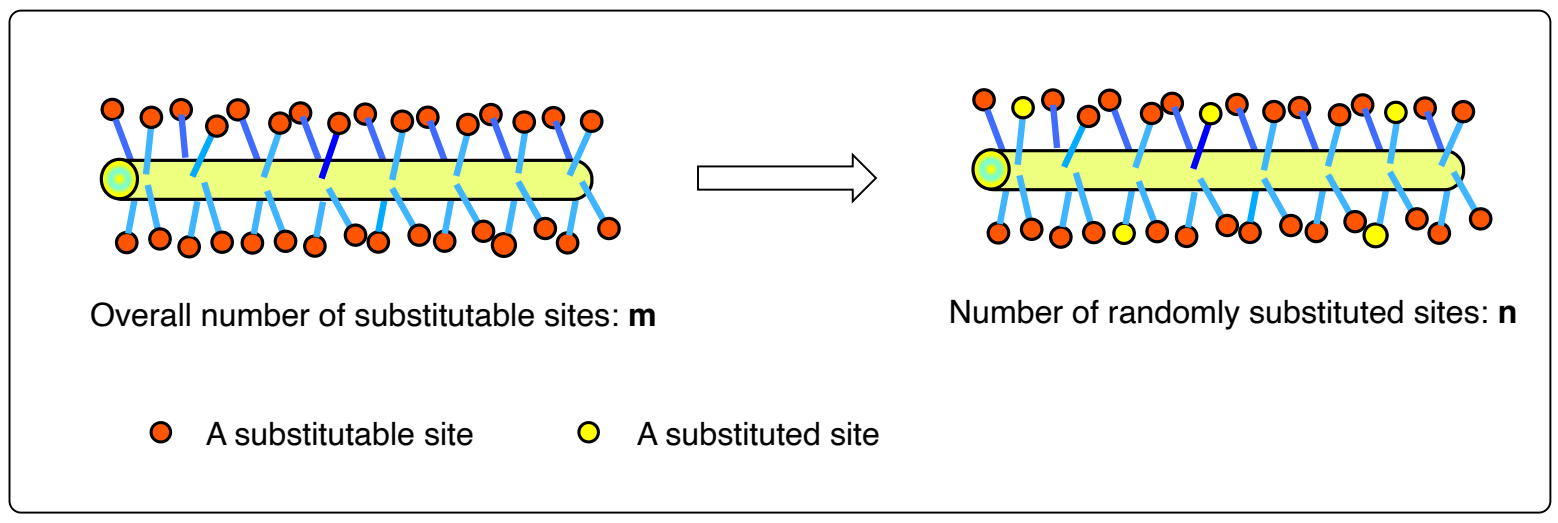

**Figure 3.** The process of partial random substitution of a parent molecule containing ***m*** substitutable sites by ***n*** substituent groups.

If the overall number of substitutable sites in the parent molecule is $m$, the number of substituted sites is $n$, the number of possible substitute groups in each randomly substituted site is $s$, then the overall number of potential isomers from this structural space $r$ can be calculated as:

$$r = C_m^n \times s^{\mathrm{n}}$$

In model system I, we choose $m$ = 1000, $s$ = 1, implying that the parent molecule contains 1000 substitutable sites, a few of these sites are randomly substituted by another group ($R^0 \rightarrow R^1$), the overall number of potential isomers can be calculated as:

$$r = C_{1000}^n$$

the calculated results are listed in Table 1.

**Table 1.** Overall number of potential isomers of model system I.

| *n* | overall substitute rate（%） | Overall number of potential isomers |
|---|---|---|
| 0 | 0 | 1 |
| 1 | 0.1 | 1000 |
| 2 | 0.2 | $4.995\times10^{5}$ |
| 3 | 0.3 | $1.66\times10^{8}$ |
| 4 | 0.4 | $4.14\times10^{10}$ |
| 5 | 0.5 | $8.25\times10^{12}$ |
| 6 | 0.6 | $1.37\times10^{15}$ |
| 7 | 0.7 | $1.94\times10^{17}$ |
| 8 | 0.8 | $2.41\times10^{19}$ |
| 9 | 0.9 | $2.66\times10^{21}$ |
| 10 | 1.0 | $2.63\times10^{23}$ |
| 11 | 1.1 | $2.37\times10^{25}$ |
| 12 | 1.2 | $1.95\times10^{27}$ |
| 13 | 1.3 | $1.48\times10^{29}$ |
| 14 | 1.4 | $1.05\times10^{31}$ |
| 15 | 1.5 | $6.88\times10^{32}$ |
| 16 | 1.6 | $4.24\times10^{34}$ |
| 17 | 1.7 | $2.45\times10^{36}$ |
| 18 | 1.8 | $1.34\times10^{38}$ |
| 19 | 1.9 | $6.92\times10^{39}$ |
| 20 | 2.0 | $3.39\times10^{41}$ |
| 21 | 2.1 | $1.58\times10^{43}$ |
| 22 | 2.2 | $7.05\times10^{44}$ |
| 23 | 2.3 | $3.00\times10^{46}$ |
| 24 | 2.4 | $1.22\times10^{48}$ |
| 25 | 2.5 | $4.76\times10^{49}$ |

From Table 1 we can see that for this model system, the overall number of potential isomers increases exponentially as the number of substituted site increases. If $n = 25$, then the overall number of potential isomers of this structural space are $4.76\times10^{49}$. This number is sufficient huge for single-molecule mixtures to exist in ***submolar-scale***. Thus, we can conclude that if a polymer molecule contains 1000 substitutable sites, 2.5% of these sites are randomly substituted by another group, then the potential isomers exceed $4.76\times10^{49}$, if this substitution process proceed at sub-molar-scale, then the products are most likely existing as single-molecule mixture state!

In model system II, we choose $m = 100$, $s = 10$, implying that the parent molecule contains 100 substitutable sites, a few of these sites are randomly substituted by one of the ten possible R group [$R^0 \rightarrow (R^{1\text{ to }10})$], the overall number of potential isomers can be calculated as:

$$r = C_{100}^{n}\times10^{n}$$

the calculated results are listed in Table 2.

**Table 2.** Overall number of potential isomers of model system II.

| n | overall substitute rate（%） | Overall number of potential isomers |
| --- | --- | --- |
| 0 | 0 | 1 |
| 1 | 1 | 1000 |
| 2 | 2 | $4.95\times10^{5}$ |
| 3 | 3 | $1.62\times10^{8}$ |
| 4 | 4 | $3.92\times10^{10}$ |
| 5 | 5 | $7.53\times10^{12}$ |
| 6 | 6 | $1.19\times10^{15}$ |
| 7 | 7 | $1.60\times10^{17}$ |
| 8 | 8 | $1.86\times10^{19}$ |
| 9 | 9 | $1.90\times10^{21}$ |
| 10 | 10 | $1.73\times10^{23}$ |

| 11 | 11 | $1.41\times10^{25}$ |
|---|---|---|
| 12 | 12 | $1.05\times10^{27}$ |
| 13 | 13 | $7.11\times10^{28}$ |
| 14 | 14 | $4.42\times10^{30}$ |
| 15 | 15 | $2.53\times10^{32}$ |
| 16 | 16 | $1.35\times10^{34}$ |
| 17 | 17 | $6.65\times10^{35}$ |
| 18 | 18 | $3.07\times10^{37}$ |
| 19 | 19 | $1.32\times10^{39}$ |
| 20 | 20 | $5.36\times10^{40}$ |
| 21 | 21 | $2.04\times10^{42}$ |
| 22 | 22 | $7.33\times10^{43}$ |
| 23 | 23 | $2.49\times10^{45}$ |
| 24 | 24 | $7.98\times10^{46}$ |
| 25 | 25 | $2.43\times10^{48}$ |

From Table 2 we can see that for this model system, as the number of substituted site increases, the overall number of potential isomers also increases exponentially. If $n = 25$, then the overall number of potential isomers of this structural space is $2.43\times10^{48}$. This number is also sufficient huge for single-molecule mixtures to exist in ***submolar-scale***. Thus, if this substitution process proceeds at sub-molar-scale, then the products are also most likely existing as single-molecule mixture state.

Finally, to better illustrate the structural feature of single molecule mixture form polymers, a model 24 mer of *O*-propyl substituted D-mannitol system was built, as illustrated in Figure 4A[15]. This system contains 192 phenolic hydroxyl group, which were randomly substituted by *n*-propyl or *i*-propyl group. The molecular weight of this system is 23775.1700 Da, and the number of structural isomers is $2^{192}$, namely, $6.277\times10^{57}$. This number, as calculated above, is sufficient huge for single-molecule mixtures to exist in even **multi-ton scales**.

**Figure 4.** (A) A model 24 mer of *O*-propyl substituted D-mannitol system. (B) Hybrid molecular structure of the model single molecule mixture system.

In this model system, although the structure of each molecules is different, in that the R group in each substitute site varied, they come randomly from the same structure space, inspired by the concept of orbital hybridization[16-19], if we define Pr* as a hybrid structure of *n*-Pr and *i*-Pr, then this single-molecule mixture system can be viewed as a "pure substance system", in that each phenolic hydroxyl group is substituted by an unprecedented hybridized Pr* group, as illustrated in Figure 4B. Thus, it is expected that the physical proprieties of this single-molecule mixture system will to some extent like a pure substance system.

In summary, single-molecule mixture, the opposite extreme form of molecules compared to "absolute pure" form, have been theoretically studied via a combination of model construction and mathematical analysis, and a series of interesting results were obtained from this study. For example, from a model polymer system calculation it is concluded that if a polymer molecule contains 1000 substitutable sites, 2.5% of these sites are randomly substituted by another group, then the potential isomers exceed $4.76\times10^{49}$, if this substitution process proceed at sub-molar-scale, then the products are most likely existing as single-molecule mixture state. These results provide theoretical evidence that single-molecule mixture state may indeed exist in realistic synthetic or natural polymer system. It is hoped that this study will inspire further exploration of this intriguing area.

## ASSOCIATED CONTENT

### Supporting Information

The Supporting Information is available free of charge on the ACS Publications website. (PDF)

## AUTHOR INFORMATION

### Corresponding Authors

**Yu Tang —** State Key Laboratory of Bioorganic and Natural Products Chemistry, Center for Excellence in Molecular Synthesis, Shanghai Institute of Organic Chemistry, University of Chinese Academy of Sciences, Chinese Academy of Sciences, Shanghai 200032, China; orcid.org/0000-0002-4272-2234; Email: tangyu@sioc.ac.cn

### Notes

The authors declare no competing financial interest.

## ACKNOWLEDGMENT

Financial support from the Youth Innovation Promotion Association of CAS (2021251) is acknowledged.

**TOC.**

**Two extreme forms of molecules**

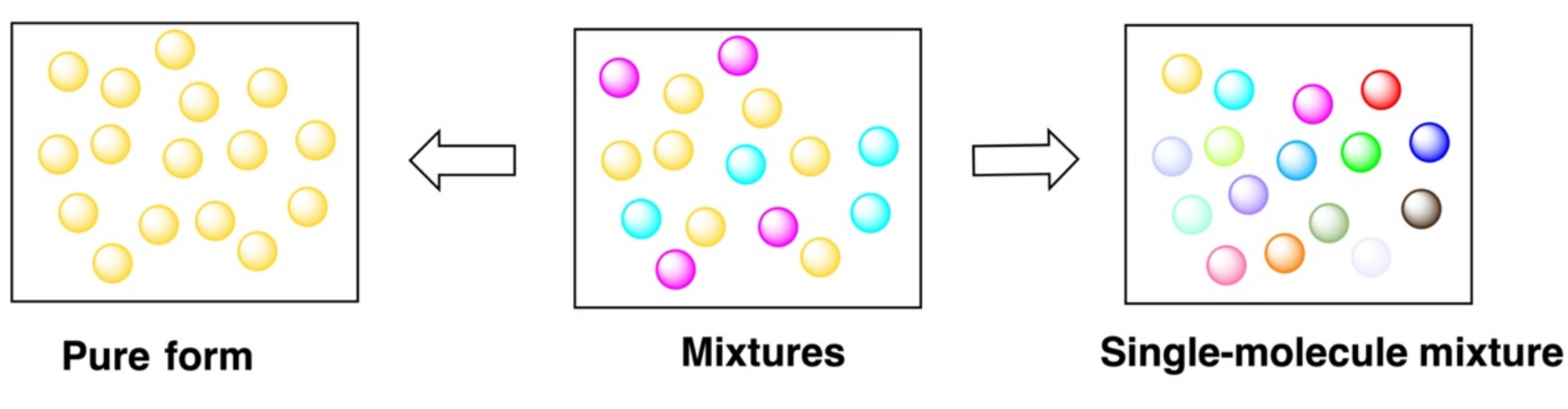

**a molecule**